\documentclass[amsmath,showpacs,amssymb,aip,jmp,floatfix,12pt]{revtex4-1}

\usepackage{graphicx}
\usepackage{dcolumn}
\usepackage{bm}
\usepackage{epsfig}
\usepackage{subfigure}
\usepackage{multirow}
\usepackage{rotating}
\usepackage{natbib}
\usepackage{amssymb}
\usepackage{graphicx}
\usepackage{textcomp}

\newtheorem{thm}{Theorem}

\begin{document}

\title{Using symmetry to generate solutions to the Helmholtz equation inside an equilateral triangle}

\author{Nathaniel Stambaugh}
 \email{nstambaugh@flsouthern.edu}
 \affiliation{Department of Computer Science and Mathematics,\\ Florida Southern College, Lakeland, Florida. 33801, USA}
\author{Mark Semon}%
 \affiliation{Department of Physics and Astronomy,\\ Bates College, Lewiston, Maine. 04240, USA}%

\date{\today}

\begin{abstract}
We prove that every solution of the Helmholtz equation $\nabla^2 \psi + k^2 \psi = 0$ within an equilateral triangle, which obeys the Dirichlet conditions on the boundary, is a member of one of four symmetry classes.  We then show how solutions with different symmetries, or different values of $k^2$, can be generated from any given solution using symmetry operators or a differential operator derived from symmetry considerations.  Our method also provides a novel way of generating the ground state solution (that is, the solution with the lowest value of $k^2$). Finally, we establish a correspondence between solutions in the equilateral and $(30^{\circ},60^{\circ}, 90^{\circ})$ triangles. \end{abstract}



\maketitle

\section{Introduction.}\label{sec:intro1}

The solutions to many important physical problems, such as electromagnetic waves in waveguides \cite{Liboff}, lasing modes in nanostructures \cite{Chang}, the electronic structure of graphene \cite{Kaufman} and the quantum eigenvalues and eigenfunctions for various potential energies  \cite{Doncheski} are obtained by solving the ubiquitous Helmholtz equation

\begin{equation}
\nabla^2 \psi + k^2 \psi = 0. \label{helmholtz}
\end{equation}

In this paper we discuss the solutions to this equation when the region of interest is an equilateral triangle ($\Delta$) and when the solutions vanish on the boundary (i.e. when they satisfy the Dirichlet condition $\psi\big|_{\partial \Delta} = 0$). Although the explicit solutions in this case are well-known, (\cite{Chang}, \cite{Doncheski}),\cite{Itzykson}, \cite{McCartin}) we present an alternative method of obtaining them that does not involve solving the differential equation directly, but rather uses only symmetry arguments, or a differential operator derived from symmetry considerations alone, to generate new solutions from any given solution.  Our method is based upon first showing that each solution within the equilateral triangle is a member of one of four symmetry classes, and then introducing symmetry operators and a differential operator which transform solutions in one symmetry class into those of another or from one value of $k^2$ to another.

Obviously any method that generates one or more new solutions to the Helmholtz equation from a given solution is quite a powerful and useful tool.  The fact that the new solutions are generated from symmetry transformations alone rather than by solving the Helmholtz equation directly makes the method even more attractive. The method also has the advantage of being able to produce solutions with prescribed symmetries, which can be important if the desired solution needs, or is known to possess, certain symmetries.  

The paper is structured as follows: in Section \ref{math} we establish our notation while reviewing the results from representation theory and linear algebra which are used in the rest of the paper.  In Section \ref{sec:rep3} we show that every solution to the Helmholtz equation within an equilateral triangle, which obeys the Dirichlet conditions on the boundary, is a member of one of four symmetry classes. In Section \ref{sec:res1} we show how to take a solution from any one of the four classes and generate from it solutions in a different symmetry class and/or with different values of the scalar $k^2$.  In Section \ref{sec:ground} we use our approach to obtain the explicit solution to the Helmholtz equation with the lowest value of $k^2$, i.e. the ``ground state solution."  In Section \ref{sec:con1} we summarize our results and discuss the various ways in which they can be applied. In particular, we discuss the correspondence between solutions in the equilateral and $(30^{\circ},60^{\circ}, 90^{\circ})$ triangles. 
\section{Notation and Background} \label{math}

\subsection{Representation Theory} \label{reps}

A representation is a homomorphism $\rho$ from the group $G$ into the group of linear transformations of a vector space (in our case, the real numbers suffice), which we denote by $\rho: G \rightarrow $GL$_n(\mathbb{R})$.  The representation assigns to each group element a transformation of the vector space that is consistent with the multiplication table of the group. For the dihedral group $\mathcal{D}_3$, every such representation can be decomposed into a direct product of three irreducible representations. 

\begin{figure}
\centering
\epsfig{file=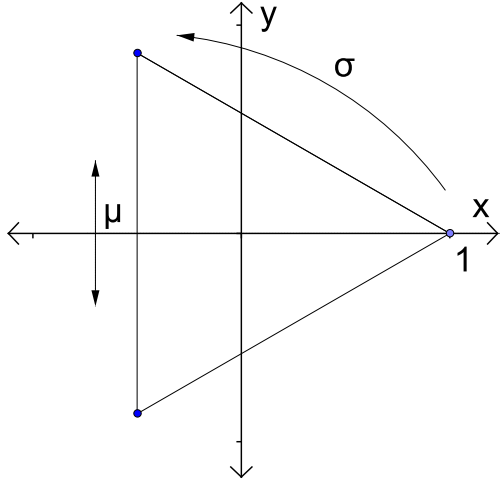,width=.4\textwidth}
\caption{The generators of the symmetry group for the equilateral triangle. We let $\sigma$ denote a counter-clockwise rotation by $120^{\circ}$ and $\mu$ the reflection in the $x$-axis.}\label{group}
\end{figure}

Before we can describe these homomorphisms we need to describe the elements of the group $\mathcal{D}_3$. Let $\sigma$ be a $120^{\circ}$ counter-clockwise rotation about the center of an equilateral triangle and $\mu$ be a reflection (without loss of generality) about the $x-$axis, as shown in Figure (\ref{group}). The defining relationship of the dihedral group says that $\mu \sigma = \sigma^{-1} \mu$. Since these two elements generate the whole group, we need only define each homomorphism on these generators. Listing the elements of the group, we have 

$$\mathcal{D}_3 = \{e, \sigma, \sigma^2, \mu, \mu\sigma, \mu \sigma^2\}.$$

The first irreducible representation is called the \textit{trivial} representation because it maps every group element to the identity map of $\mathbb{R}$. While this may seem somewhat, well, trivial, it actually plays an interesting role later on. Symbolically, $\rho_1(\alpha) = 1$ for every $\alpha \in \mathcal{D}_3$. 

The second representation is called the {\it sign} representation, though it also could be called the \textit{orientation} representation, because it shows whether a reflection has occurred. That is, $\rho_2(\sigma) = 1$ and $\rho_2(\mu) = -1$. Obviously the trivial and sign representations are one dimensional.

The third representation is the only one that displays every nuance in the group, and it therefore is sometimes used to define $\mathcal{D}_3$. Unlike the two previous representations, $\rho_3$ is a two dimensional representation whose elements (in GL$_2(\mathbb{R})$) are
\begin{center}
\begin{tabular}{ccc}
$\rho_3(\sigma) = \frac{1}{2} \begin{pmatrix} -1 & -\sqrt{3} \\ \sqrt{3} & -1 \end{pmatrix}$, & \hspace{1mm} $\rho_3(\mu) = \begin{pmatrix} 1 & 0 \\ 0 & -1 \end{pmatrix}$.
\end{tabular}
\end{center}

Using $\rho_3$ we can define in a natural way the action of each group element $\alpha$ on a solution $f(x,y)$ of the Helmholtz equation . The argument of the function is the vector $\left[\begin{tabular}{c} $x$ \\ $y$ \end{tabular} \right]$ in $\mathbb{R}^2$, so the following action is well defined:
\begin{equation}
(\alpha \cdot f)(x,y) = f\Big(\rho_3(\alpha)^{-1} \left[\begin{tabular}{c} $x$ \\ $y$ \end{tabular} \right]\Big). \label{action}
\end{equation}
The use of the inverse of the representation matrix is required to make this a homomorphism, and can be thought of as a passive transformation on the coordinates. In order to simplify the notation, we will write $\alpha f$ in place of $\alpha\cdot f$.

\subsection{Inner Product Spaces}\label{innerprod}
If $f_1$ and $f_2$ are two solutions of the Helmholtz equation then we define the inner product of $f_1$ and $f_2$ as

\begin{equation}
\langle f_1, f_2 \rangle = \int\int_{\Delta} f_1(x,y)f_2(x,y)dxdy,
\end{equation}
where the integral is taken over the domain $\Delta$. The norm (or length) of a solution $f$ is defined as  
\begin{equation}
||f|| = \sqrt{\langle f,f \rangle}.
\end{equation}
This is called the $\mathcal{L}^2$ norm, which we will use to normalize any given solution and also to establish when two solutions $f_1$ and  $f_2$ are orthogonal ($\langle f_1, f_2 \rangle =0 \Longleftrightarrow f_1 \perp f_2$). 

If two solutions have the same value of $k$ and are orthogonal, we can form a two dimensional space spanned by these solutions. In the same way that we use the vector $\left[\begin{tabular}{c} $a$ \\ $b$ \end{tabular} \right]$ to represent $a \hat{i} + b \hat{j}$, in our context this column vector will represent $a f_1 + b f_2$.

\section{\label{sec:rep3} Classifying solutions to the Helmholtz equation by their symmetries.}
If $f$ is a solution of the Helmholtz equation within a region (denoted by $\Delta$) whose sides form an equilateral triangle, and which satisfies the Dirichlet conditions on the boundary, then for each element $\alpha \in D_3$, $\alpha  f$ (as defined by Eq. (\ref{action})) is a solution of the Helmholtz equation with the same value of $k^2$. In this section we prove that every such solution $f$ belongs to one of four sets according to its rotational and reflection symmetries.  We call these sets {\it symmetry classes} and denote them by A1, A2, E1 and E2.

We first consider the rotational symmetries of solutions of the Helmholtz equation.  If a solution $f$ is rotated to obtain a new solution $\sigma  f$ then, in general, the new solution can be rotationally symmetric ($\sigma  f = f$), rotationally anti-symmetric ($\sigma  f = -f$), or rotationally asymmetric ($\sigma  f \neq \pm f$). We can eliminate the rotationally anti-symmetric case as follows: Suppose that $\sigma  f = - f$.  Then, since $\sigma^3$ is the identity element, 

\begin{equation}
f = \sigma^3  f =  \sigma  \sigma  (- f) = \sigma  f = -f.
\end{equation}

\noindent Thus $f(x) = -f(x)$ for every $x\in \Delta$ so $f$ is identically zero on the domain.  Therefore, when we rotate a (non-trivial) solution $f$ to obtain a new solution $\sigma f$, the new solution $\sigma f$ must be either rotationally symmetric or rotationally asymmetric.  In what follows we first consider the effect of reflections on the rotationally symmetric solutions and then the effect of reflections on the rotationally asymmetric solutions.

\subsection{Properties of rotationally symmetric solutions under reflection.} \label{rotsym1}

Assume that $f$ is a rotationally symmetric solution of the Helmholtz equation, so $\sigma  f = f$. In general, the solution $\mu f$ can be symmetric, anti-symmetric, or asymmetric under reflection. However, we can eliminate the asymmetric case as follows: Suppose $\mu  f \neq \pm f$.  Define the two functions
\begin{eqnarray}
f_+ & = & \frac{1}{2}(f + \mu  f),\\
f_- & = & \frac{1}{2}(f - \mu  f).
\end{eqnarray}

The functions $f_+$ and $f_-$ are solutions of the Helmholtz equation because each is constructed from a linear combination of solutions to the Helmholtz equation. Furthermore, $f_+$ is symmetric and $f_-$ is anti-symmetric under reflections about the $x-$axis, and the solution $f$ can be written as $f=f_+ + f_-$. In addition, the boundary condition obeyed by $f$ will also be obeyed by $f_+$ and $f_-$. Consequently, any asymmetric solution $f$ can be decomposed into the sum of the symmetric and anti-symmetric solutions $f_+$ and $f_-$, and we thus need only consider solutions to the Helmholtz equation which are symmetric or anti-symmetric under reflection about the x-axis. 

\begin{figure}
\centering
\subfigure[\, A1\label{A1}]{\epsfig{file=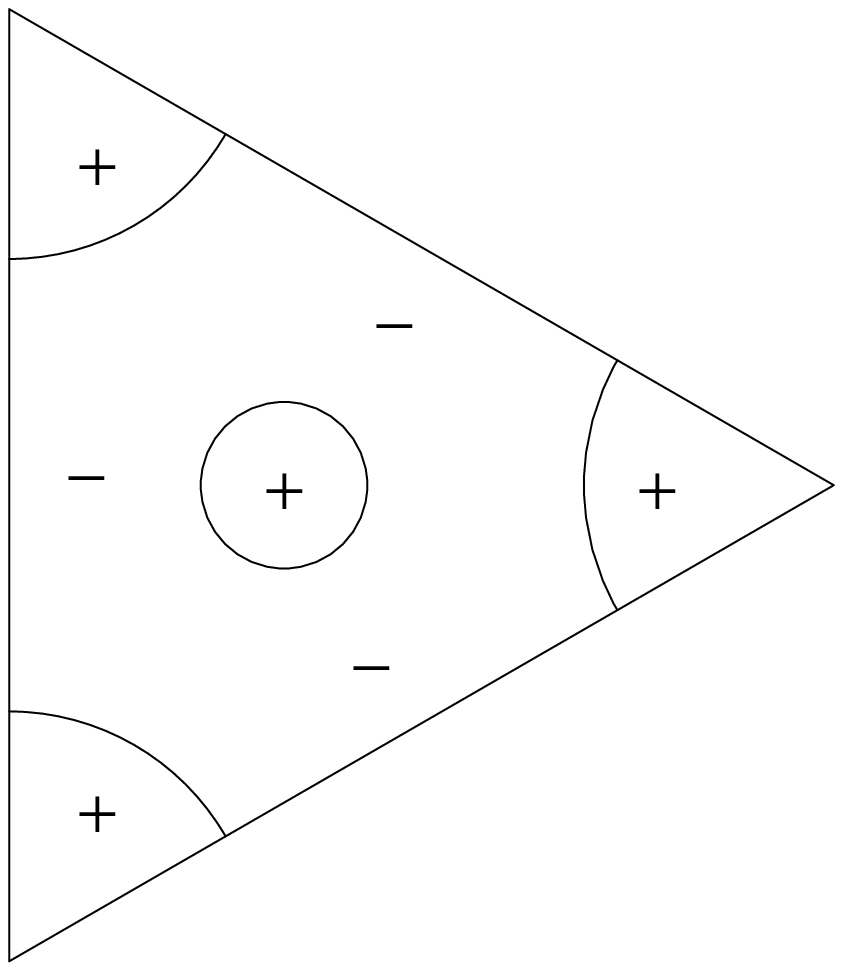,width=.2\textwidth}}\hspace{8mm}
\subfigure[\, A2\label{A2}]{\epsfig{file=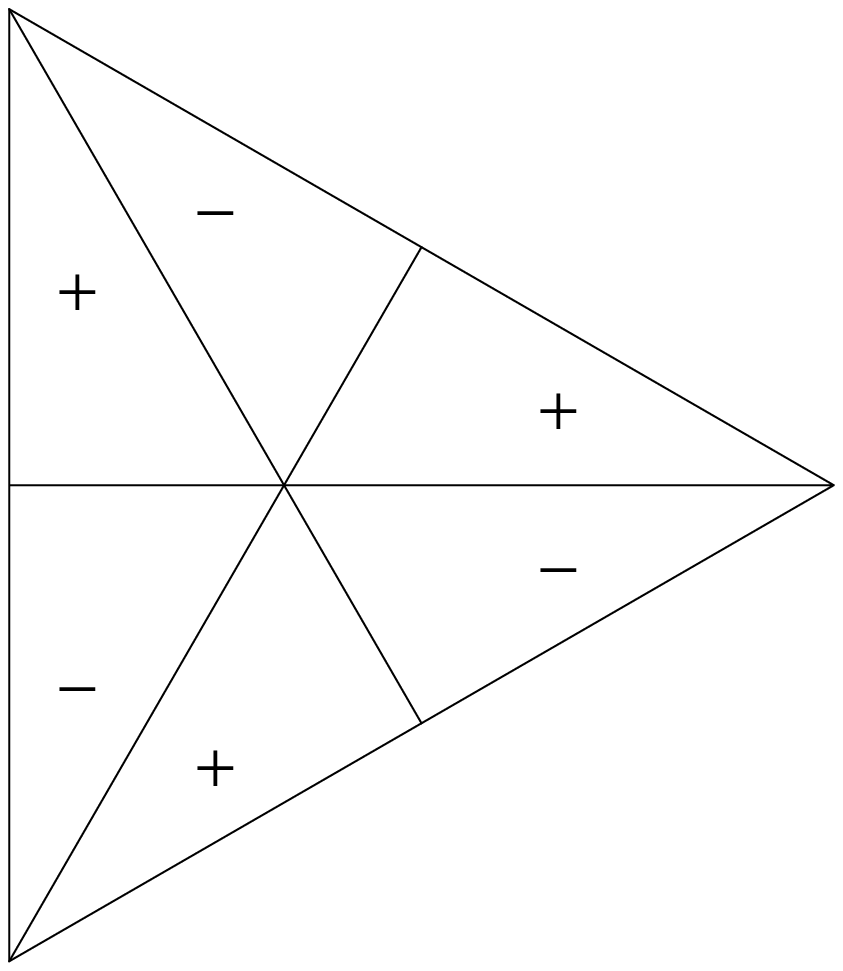,width=.2\textwidth}}
\caption{Figures (a) and (b) show solutions $f$ which both have rotational symmetry. The solution shown in (a) is also symmetric under reflection, while the solution shown in (b) is also anti-symmetric under reflections.}
\label{sym}
\end{figure}

If we denote by A1 the set of solutions that are symmetric under a rotation $\sigma$ and symmetric under a reflection $\mu$, and by A2 the set of solutions that is symmetric under a rotation $\sigma$ and anti-symmetric under reflection $\mu$ , then we can summarize the results of this section with the following table: 

\begin{center}
\begin{table}[h]
\caption{Rotationally Symmetric Solutions}
\begin{tabular}{c||c|c}
&  \; $\sigma  f_i$ \; & \; $\mu  f_i$ \;  \\ \hline \hline
$f_1\in$A1 & $+f_1$ & $+f_1$ \\
$f_2\in$A2 & $+f_2$ & $-f_2$
\end{tabular}
\end{table}
\end{center}
Figure (\ref{sym}) shows examples of solutions in the symmetry classes A1 and A2.

It's interesting to note that all of the solutions in A1 are orthogonal to all the solutions in A2: Suppose $f_1\in A1$, then $\mu f_1 = f_1 \Longleftrightarrow f_1(x,-y) = f_1(x,y)$, which means that if $f_1\in A1$ then $f_1$ is even in $y$.  Similarly, if $f_2\in A2$, then $\mu f_2 = -f_2 \Longleftrightarrow f_2(x,-y) = -f_2(x,y)$, which means that if $f_2\in A2$ then $f_2$ is odd in $y$. Therefore the product $f_1 f_2$ is odd in $y \Rightarrow \langle f_1, f_2 \rangle =0 \Longleftrightarrow f_1 \perp f_2$

We end this subsection by noting that $f_1\in$ A1 is characterized in the trivial representation by
\begin{equation}
\alpha f_1 = \rho_1(\alpha) f_1\label{eqn:A1},
\end{equation}
and $f_2\in$A2 is characterized in the sign representation 
\begin{equation}
\alpha f_2 = \rho_2(\alpha) f_2.\label{eqn:A2}
\end{equation}

\subsection{\label{sec:rep5} Properties of rotationally asymmetric solutions under reflection.}

As was shown at the beginning of  Section \ref{sec:rep3}, if $f$ is a solution of the Helmholtz equation in the region $\Delta$ then the rotated solution $\sigma f$ will be either rotationally symmetric or rotationally asymmetric.  Having examined the reflection properties of the rotationally symmetric solutions in the previous section we now examine the reflection properties of the rotationally asymmetric solutions. Although the lack of rotational symmetry might make us expect that these solutions will be of little use, on the contrary, not only are they the most common solutions to the Helmholtz equation, but they also have a number of interesting and useful properties.

Let E1 be the set of solutions which are asymmetric under rotations and symmetric under reflections, and E2 be the set of solutions which are asymmetric under rotations and anti-symmetric under reflections.  Consider a normalized solution $f_1$ in class E1. At this point we do not know anything about $\sigma  f_1$ except that it is a solution with the same value of $k^2$ as $f_1$. For that matter, so is $\sigma^2  f_1$. So consider the function $\hat{f}_2 = \sigma  f_1 - \sigma^2  f_1$. We now show that $\hat{f}_2$ is in symmetry class E2:

\begin{eqnarray}
\mu \hat{f}_2 & = & \mu ( \sigma f_1 - \sigma^2 f_1) \nonumber \\
& = & \mu \sigma f_1 - \mu \sigma^2 f_1 \nonumber \\
& = & \sigma^2\mu f_1 - \sigma \mu f_1 \nonumber \\
& = & \sigma^2 f_1 - \sigma f_1 \nonumber \\
& = & - \hat{f}_2
\end{eqnarray}

The reason we call this new solution $\hat{f}_2$ is that it is not normalized; we will call the normalized function $f_2$. We also note that $f_1$ and $f_2$ are orthogonal since their product is odd in $y$.

Since the action of the group introduces a second solution with the same value of $k^2$, we consider the two dimensional solution space spanned by $f_1$ and $f_2$. As in Section \ref{A1}, the vector $\vec{f}=\left[ \begin{tabular}{c} $a$ \\ $b$ \end{tabular} \right]$ represents the solution $f = a f_1 + b f_2$. Written in this way, we can recognize the third irreducible representation $\rho_3$
$$\alpha \vec{f} = \rho_3(\alpha) \vec{f}.$$
This equation actually contains a lot of information, and is strikingly similar to Eqs.(\ref{eqn:A1}) and (\ref{eqn:A2}). We now use it to normalize $\hat{f}_2$.

Let $\vec{f_1} = \left[ \begin{tabular}{c} 1 \\ 0 \end{tabular} \right]$. Then $\sigma f_1$ can be re-expressed as a linear combination of $f_1$ and $f_2$ as follows:
\begin{eqnarray}
\sigma f_1 & = & \frac{1}{2} \begin{pmatrix} -1 & -\sqrt{3} \\ \sqrt{3} & -1 \end{pmatrix} \begin{pmatrix} 1 \\ 0 \end{pmatrix} \\
& = & \frac{1}{2}\begin{pmatrix} -1 \\ \sqrt{3} \end{pmatrix} \nonumber  \\
& = & - \frac{1}{2} f_1 + \frac{\sqrt{3}}{2} f_2.
\end{eqnarray}

Similarly, $\sigma^2 f_1 = - \frac{1}{2} f_1 - \frac{\sqrt{3}}{2} f_2.$  Therefore

\begin{equation}
\sigma f_1 - \sigma^2 f_1 = \sqrt{3} f_2,
\end{equation}

and solving for $f_2$, we find the normalized solution
 
\begin{equation}\label{f2}
f_2=\frac{1}{\sqrt{3}}(\sigma f_1 - \sigma^2 f_1).
\end{equation}

\begin{figure}
\centering
\subfigure[\, E1\label{E1}]{\epsfig{file=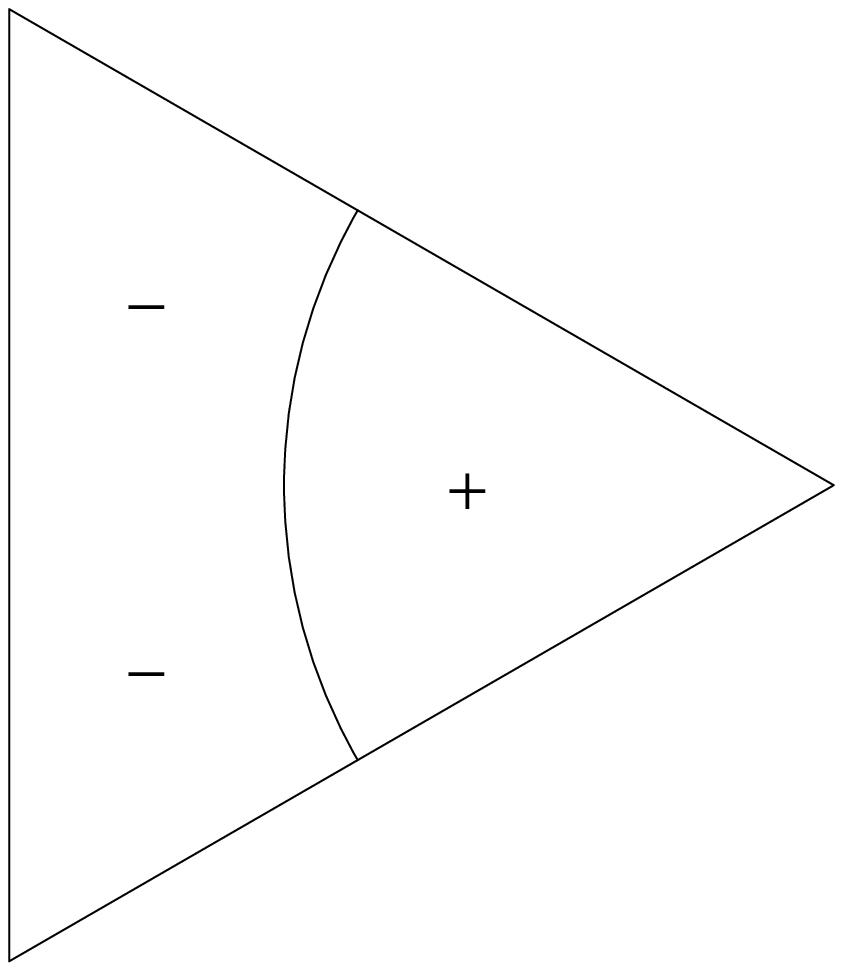,width=.2\textwidth}}\hspace{8mm}
\subfigure[\, E2\label{E2}]{\epsfig{file=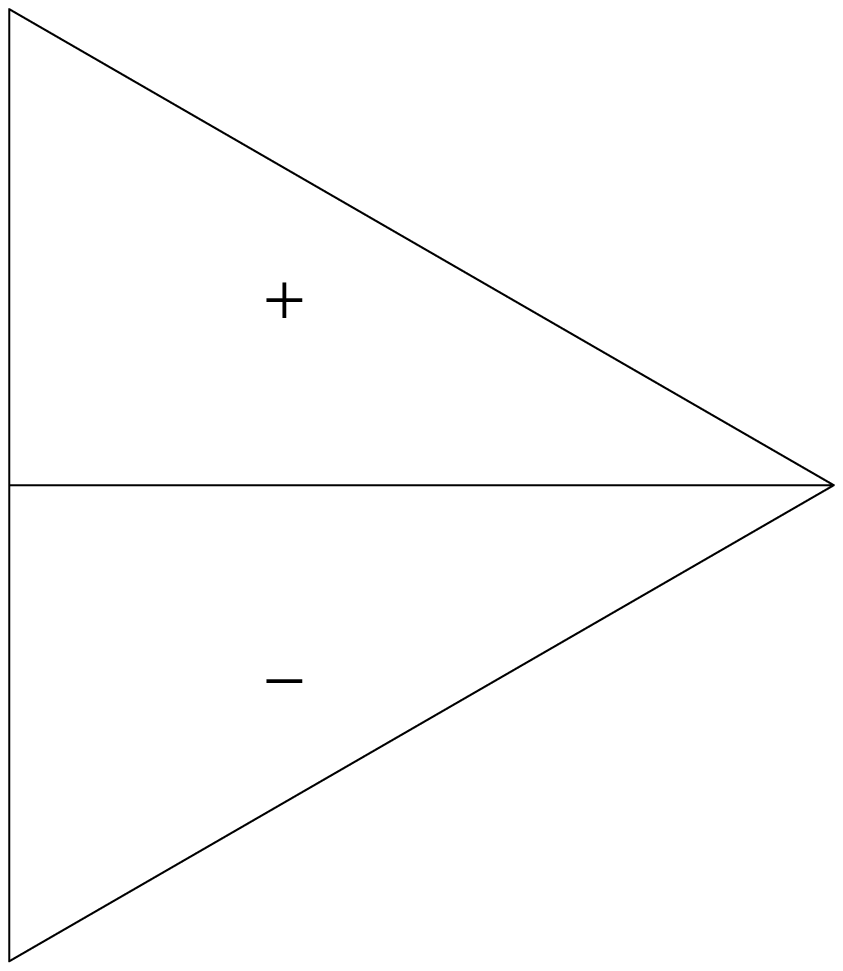,width=.2\textwidth}}
\caption{Figures (a) and (b) show two solutions $f$, neither of which has rotational symmetry. The solution $f_1$ in (a) is symmetric under reflection and in the set E1, while the solution $f_2$ in (b) is anti-symmetric under reflection and in the set E2.}
\label{sym2}
\end{figure}

Similarly, $f_1$ can be expressed in terms of $f_2$ as
\begin{equation}\label{f1}
f_1=\frac{1}{\sqrt{3}}(\sigma f_2 - \sigma^2 f_2).
\end{equation}
Consequently, given any solution $f_1$ in the symmetry class E1 we can generate from it an orthogonal solution $f_2$ in the symmetry class E2, and \textit{vice versa}.  Figu \ref{sym2} shows two examples of solutions from E1 and E2.

To summarize, in this section we have established the important result that every solution to the Helmholtz equation within an equilateral triangle can be placed into one of four symmetry classes.  The class A1 is the set of solutions which are symmetric under rotation and under reflection, the class A2 is the set of solutions which are symmetric under rotation and anti-symmetric under reflection, the class E1 is the set of solutions which are asymmetric under rotation and symmetric under reflection, and the class E2 is the set of solutions which are asymmetric under rotation and anti-symmetric under reflection.  These results are summarized in Table \ref{4sym}.

\begin{table}
\caption{The four symmetry classes}\label{4sym}
\begin{tabular}{c|c|c}
& \multicolumn{2}{c}{Rotation}\\
Reflection & Symmetric & Asymmetric \\ \hline
&& \\
 Symmetric & A1 & E1 \\
&&\\ \hline
&&\\
 Anti-Symmetric & A2 & E2 \\
&&
\end{tabular}
\end{table}

\section{\label{sec:res1} Generating new solutions from a given solution.}

In the previous section we showed how Eq.(\ref{f1}) can be used to generate a solution in the symmetry class E2 (i.e. a solution that is anti-symmetric when reflected about the $x$-axis) from any solution in the symmetry class E1 (i.e. from a solution that is symmetric when reflected about the $x$-axis) and \textit{vice versa} (using Eq.(\ref{f2})).  More specifically, if we have a solution that is even in the $y$-coordinate and asymmetric under rotations $\sigma$, then we can generate from it a solution that is odd in the $y$-coordinate and asymmetric under rotations $\sigma$, and \textit{vice versa}.  Furthermore, in each case, the generated solution is orthogonal to the original solution.  In this section we show when it is possible to take a solution from one of the four symmetry classes and generate from it an orthogonal solution in one of the other symmetry classes and/or with a different value of $k^2$. 

In this section we present three ways to generate new solutions from a given solution. First, we show how to build a many-to-one correspondence between solutions in symmetry class A2 and symmetry class E2. Next we present a way to take {\it any} solution and generate from it a solution with a larger scalar, which we will call a ``harmonic" of the original solution. The final technique introduces a differential operator which transforms a solution from symmetry class A2 into a (non-trivial) solution in symmetry class A1. Applying this same technique to a solution in symmetry class A1 yields a solution in symmetry class A2. The resulting solution in A2 may be the trivial solution, and we discuss how this fact gives new insight into the ``ground state solution." 
In this section we present three ways to generate new solutions from a given solution. First, we show how to build a many-to-one correspondence between solutions in symmetry class A2 and symmetry class E2. Next we present a way to take {\it any} solution and generate from it a solution with a larger scalar, which we will call a ``harmonic" of the original solution. The final technique introduces a differential operator which transforms a solution from symmetry class A2 into a (non-trivial) solution in symmetry class A1. Applying this same technique to a solution in symmetry class A1 yields a solution in symmetry class A2. The resulting solution in A2 may be the trivial solution, and we discuss how this fact gives new insight into the ``ground state solution." 
In this section we present three ways to generate new solutions from a given solution. First, we show how to build a many-to-one correspondence between solutions in symmetry class A2 and symmetry class E2. Next we present a way to take {\it any} solution and generate from it a solution with a larger scalar, which we will call a ``harmonic" of the original solution. The final technique introduces a differential operator which transforms a solution from symmetry class A2 into a (non-trivial) solution in symmetry class A1. Applying this same technique to a solution in symmetry class A1 yields a solution in symmetry class A2. The resulting solution in A2 may be the trivial solution, and we discuss how this fact gives new insight into the ``ground state solution." 

We begin by quoting a theorem usually attributed to Lam\'e, using the statement (and referring the reader to the proof) given by McCartin \cite{McCartin}:

\begin{thm}\label{Lame}
(Lam\'e) Suppose that $T(x,y)$ is a solution to the Helmholtz equation which can be represented by the trigonometric series
\begin{align}
T(x,y) = & \sum_{i} \left( A_i \sin(\lambda_i x + \mu_i y + \alpha_i) \right. \nonumber \\
& \left. \hspace{2cm} B_i \cos(\lambda_i x + \mu_i y + \beta_i) \right),
\end{align}
with $\lambda_i^2 + \mu_i^2 = k^2$. Then
\begin{enumerate}
\item $T(x,y)$ is antisymmetric about any line along which it vanishes;
\item $T(x,y)$ is symmetric about any line along which its normal derivative, $\frac{\partial T}{\partial \nu}$, vanishes.
\end{enumerate}
\end{thm}

Lam\'e \cite{McCartin} also proves that the solutions to the Helmholtz equation in a triangular region subject to the Dirichlet conditions can be expressed in this way and that they form a complete, orthonormal set.  Explicit expressions for these solutions are also given by Doncheski \textit{et al.} \cite{Doncheski}

\subsection{\label{sec:E2_A2} E2 $\leftrightarrow$ A2} 
\begin{figure}
\centering
\epsfig{file=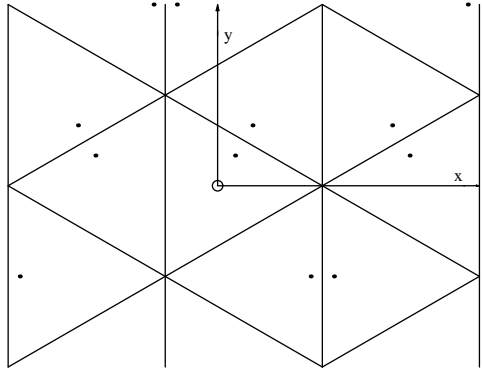,width=.4\textwidth}
\caption{\label{tile}This shows how the fundamental domain tiles the plane via reflections. Note how the location of the point in the fundamental domain changes as the reflections are made.}
\end{figure}

In order to relate solutions in the symmetry classes E2 and A2 we use a method called ``tessellating the plane,"  which extends any solution within the triangular domain to the plane.  We begin the tessellation by defining the triangular region in which we are working as the ``fundamental domain" and then we reflect this domain across each of its three boundaries. An example of this construction is shown in Figure (\ref{tile}). Tessellating the plane provides a way to smoothly extend the solution from the triangular region of interest to the plane.

\begin{figure}
\subfigure[\,E2]{\epsfig{file=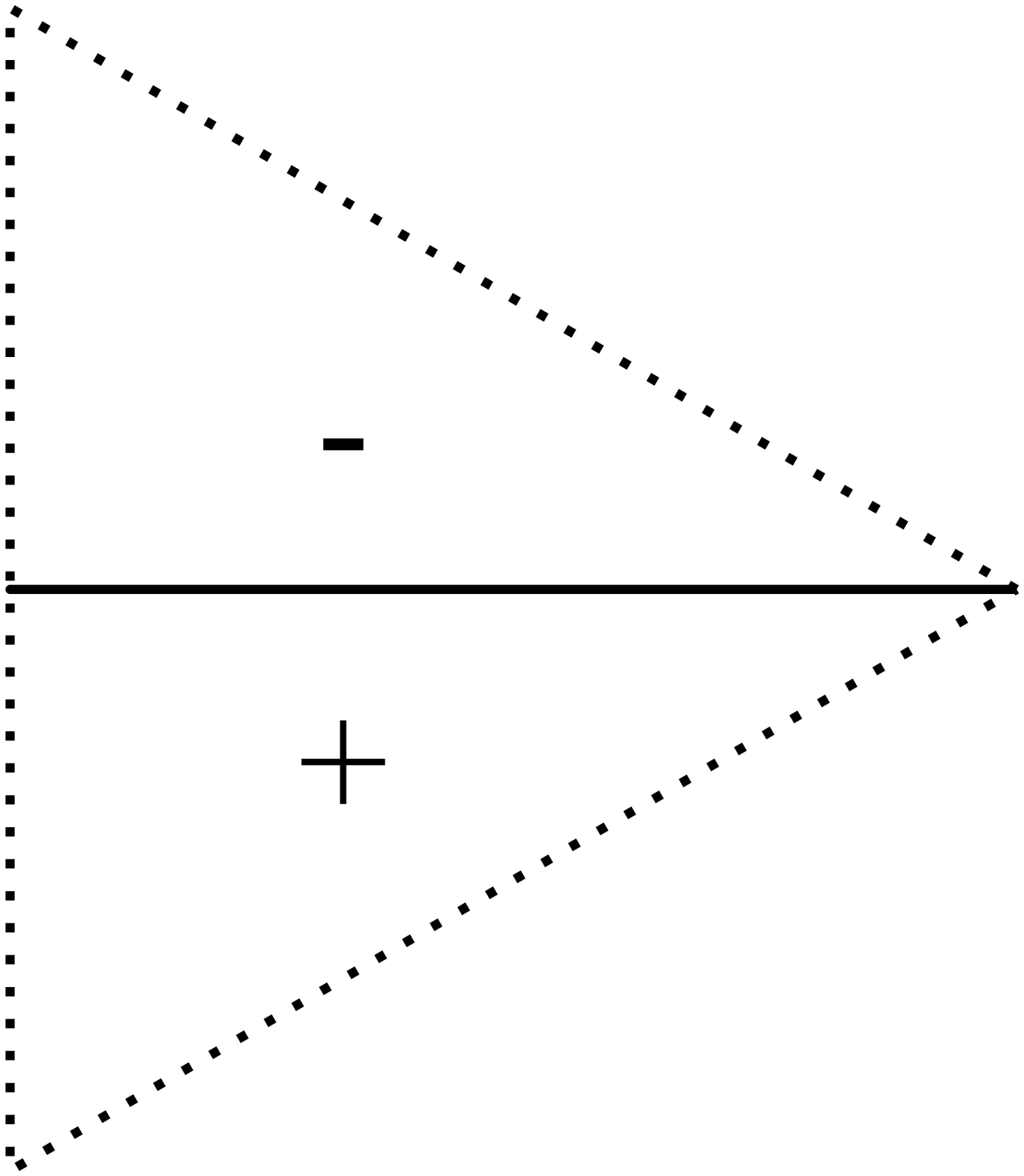,width=.15\textwidth}}
\subfigure[\,A2]{\epsfig{file=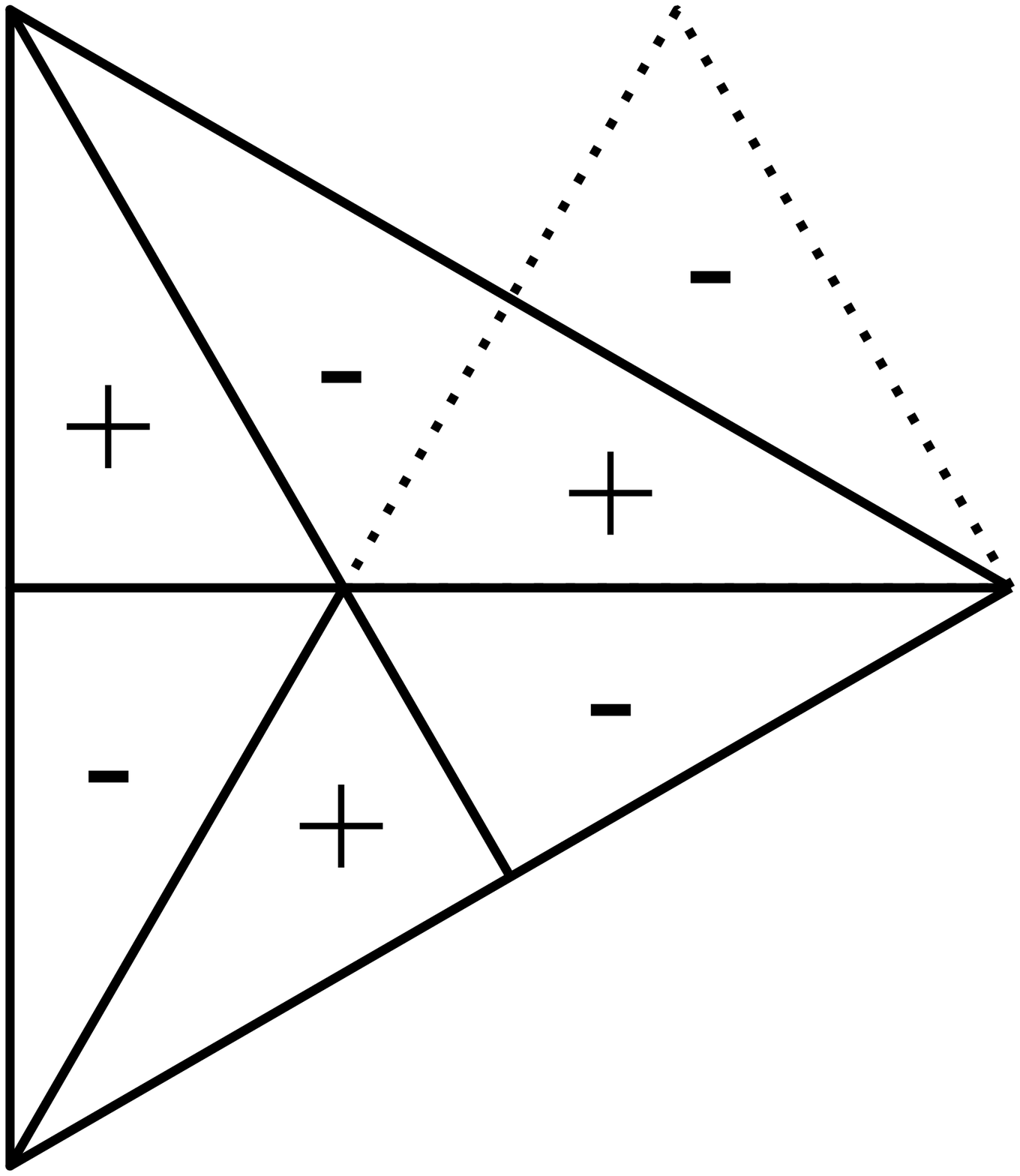,width=.15\textwidth}}
\caption{A pictoral representation of solutions in the symmetry classes E2 and A2.}
\label{E2A2}
\end{figure}

We begin by considering a solution $f_2$ in the symmetry class E2. Since $f_2$ is antisymmetric under reflection it must be zero along the $x$ axis.  Thus, any solution in the symmetry class E2 will have the form shown in Figure (\ref{E2A2}a), with possibly more nodal curves. Similarly, if a solution is in the symmetry class A2, it not only needs a nodal line along the $x$-axis, but also along the other altitudes. Thus, any solution in the symmetry class A2 will have the form shown in Figure (\ref{E2A2}b), with possibly more nodal curves. 

By examining the equilateral triangles in Figure (\ref{E2A2}) we see that the linear transformation

\begin{equation}\label{Trans}
T \left( \left[\begin{tabular}{c} $x$ \\ $y$ \end{tabular}\right] \right)= \frac{1}{6} \left[ \begin{tabular}{c} $3 x +\sqrt{3} y +3$ \\ $-\sqrt{3} x + 3 y + \sqrt{3}$ \end{tabular} \right]
\end{equation}
maps the dashed region in Figure (\ref{E2A2}a) into the dashed region in Figure (\ref{E2A2}b) [The origin is one unit from each vertex]. The A2 solution can then be constructed by tessellating the dashed triangle in Figure (\ref{E2A2}b) and thus creating the triangle in that figure who boundaries are shown with the solid lines.  Therefore, the transformation in Eq. (\ref{Trans}), which is deduced from the figures, takes {\it any} solution in the E2 symmetry class and transforms it into a solution in the A2 symmetry class.  Having determined the explicit form of the transformation from the tessellation we can rewrite the Helmholtz equation in the new coordinates and deduce the new value of the scalar term.  In this way we can construct an explicit solution in A2 from any given solution in E2 and determine the scalar term associated with it.  We carry out this calculation explicitly in Appendix I and show that if the original function in E2 is a solution of the Helmholtz equation with the scalar $k^2$, then the solution in A2 generated from it will be a solution of the Helmholtz equation with the scalar $3k^2$. Similarly, if we start with a solution in A2 (rather than E2) of the Helmholtz equation with the scalar $k^2$, then the same argument will produce another solution in A2 but with the scalar $3{k^2}$.  In summary if we start with any function in A2 or E2 that is a solution to the Helmholtz equations with the scalar $k^2$ then, using the method presented in Appendix I, we can generate a function in A2 that is a solution to the Helmholtz equation with the scalar $3k^2$.

The proof given in Appendix I is reversible, so if we start at the end of the argument with a function in A2 which is a solution of the Helmholtz equation with scalar $k^2$, then we can run the argument backwards and generate another solution which is in either A2 or E2.  In both cases the function produced will be a solution of the Helmholtz equation with the scalar \textbf{$\frac{k^2}{3}$}.  However, since there exists a solution with the lowest allowed value of $k^2$ (the ``ground state solution") the process of generating solutions with smaller values of the scalar must stop.  We discuss this situation in the next section. Note that this method establishes a many-to-one correspondence between solutions in symmetry class A2 and symmetry class E2.

\subsection{\label{sec:A1_A1} Generating solutions with higher values of $k^2$}
\begin{figure}
\centering
\subfigure[$n=2$]{\epsfig{file=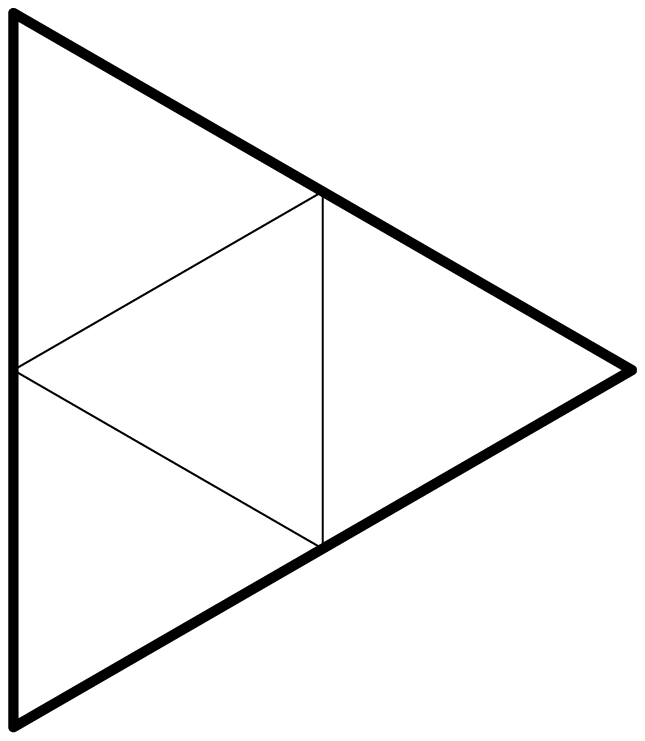,width=.2\textwidth}}\hspace{8mm}
\subfigure[$n=3$]{\epsfig{file=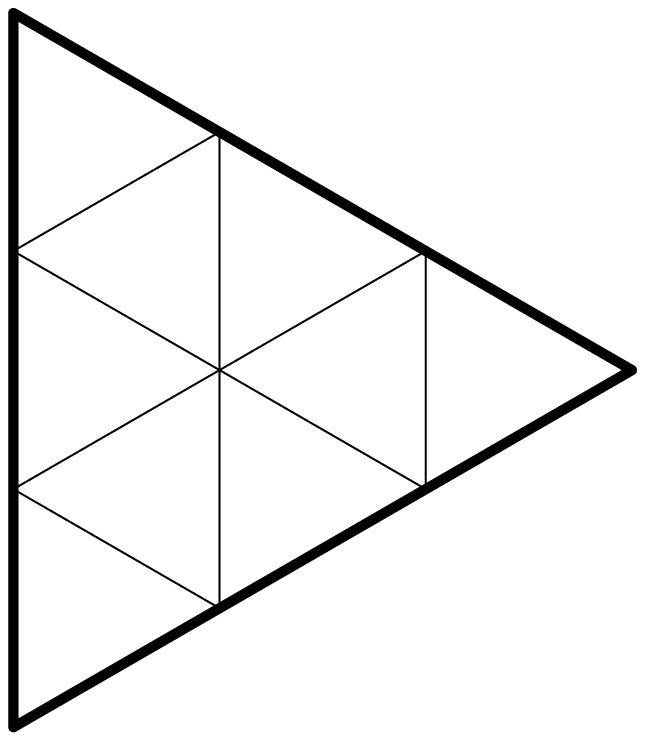,width=.2\textwidth}}
\caption{The equilateral triangle can be decomposed into $n^2$ equilateral triangles}
\label{boost}
\end{figure}
\begin{figure}
\centering
\subfigure[Two successive constructions from Section \ref{sec:E2_A2}]{\epsfig{file=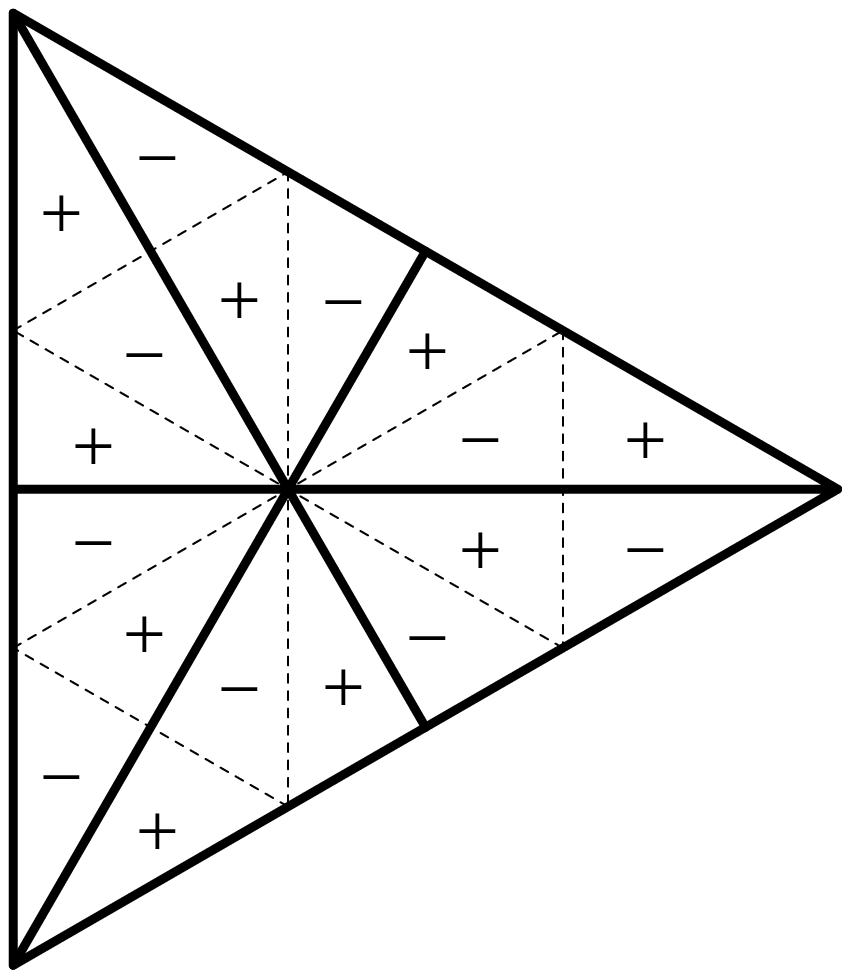,width=.2\textwidth}}\hspace{8mm}
\subfigure[Harmonic with $n=3$\label{a9}]{\epsfig{file=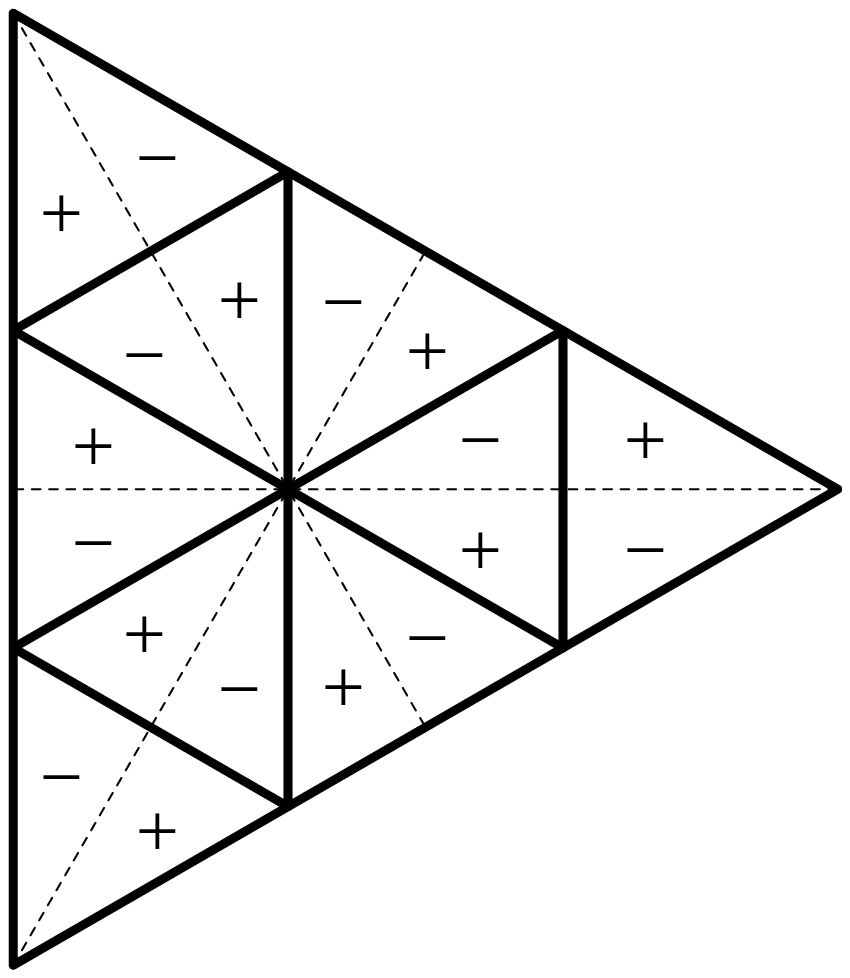,width=.2\textwidth}}
\caption{These pictures show two redundant constructions of a higher energy state.}
\label{bst_vs}
\end{figure}

Although we can use the proof presented in the Appendix I to generate from a solution in A2 with the scalar $k^2$ new solutions in A2 or E2 with three times the value of the scalar $k^2$, there is a more direct way to take a solution and generate from it solutions with higher values of the scalar for solutions from {\it any} symmetry class.  To do this we carry out a different tessellation of the plane and then extract the new value of the scalar from the coordinate transformation.  The tessellation, which is shown in Figure (\ref{bst_vs}), decomposes the equilateral triangle into $n^2$ equilateral sub-triangles, for any $n \in \mathbb{N}$. As we show in Appendix II, the explicit coordinate transformation that creates the sub-triangles is constructed from a dilation followed by a translation along the $x$-axis.   Carrying out this transformation we can start with a function in any symmetry class that is a solution of the Helmholtz equation with the scalar $k^2$ and generate from it a family of new solutions to the Helmholtz equation with scalar $n^2k^2$. Note that if the original solution is in the symmetry class E2 (i.e. is rotationally asymmetric) then this approach will produce a solution in A2 (i.e. the generated solution will become rotationally symmetric) whenever $n$ is divisible by $3$. In this case, however, we have would have already constructed this solution by using the method from Section (\ref{sec:E2_A2}) twice. See Figure \ref{bst_vs} for a schematic proof of this fact.

Note that any solution with a vertical nodal line can be reduced using this method. Extending the solution to the plane, we can use Theorem \ref{Lame} and any vertical nodal line to introduce a new mirror symmetry. Combining this with the mirror symmetries about the boundaries and the rotational symmetry, there will be a fundamental domain without vertical nodal lines. This represents a solution which can be used to generate the original solution without any vertical nodal lines.

\subsection{\label{sec:A2_A1} Generating even solutions from odd solutions and {\it vice versa}.}

In this section we prove that there exists a differential operator that transforms a solution $f_2$ in A2, i.e. a solution that is symmetric under rotations and an odd function of y, into a solution $f_1$ in A1, that is, into a solution which is symmetric under rotations and an even function of $y$.  To show this we construct a function $\hat{f_1}$ in A1 (where $f_1$ is the normalized solution) by defining $\hat{f_1}$ in the following way:

\begin{align}
	\hat{f}_1 &= \left(\frac{\partial^3}{\partial y^3} - 3 \frac{\partial^3}{\partial y\partial x^2}\right) f_2(x,y) \label{even_odd}.
\end{align}

The proof that $\hat{f_1}$ is in A1 goes as follows: First, it is easy to see that $\hat{f_1}$ is a solution to the Helmholtz equation with the scalar $k^2$ \textit{iff} $f_2$ is a solution with the same scalar $k^2$.  To complete the proof we need to show four more things: first, that $\hat{f}_1$ is rotationally symmetric, second, that $\hat{f}_1$ satisfies the Dirichlet conditions, third, that $\hat{f_1}$ is even in the $y$-coordinate, and fourth, that $\hat{f}_1$ is not identically zero.

With single variable functions, one common way of generating an even function from an odd function is to take the first derivative.  However, the need to satisfy all of the above conditions requires a more complicated procedure. The generalization of the first derivative in one dimension to two dimensions is the directional derivative, and the directional derivative of a function $f$ is $\nabla f\cdot \hat{e}$.  We can represent the directional derivative operator in the $\hat{e}$ direction as $\nabla \cdot \hat{e}$. Higher order directional derivatives are simply powers of this operator. As can be easily verified, although the first directional derivative of $f_2$ satisfies the Helmholtz equation it would not necessarily have the correct rotational symmetry. To correct this, we can {\it symmetrize} the solution by adding it to both of its rotates. The resulting solution will now be even and rotationally symmetric. Additionally, the transformed solution will satisfy the boundary (Dirichlet) condition since any nodal line parallel to a side will remain a nodal line. This follows from Theorem \ref{Lame}, since every solution is antisymmetric in a nodal line. Thus a directional derivative along the nodal line is zero, and by anti-symmetry the other two directional derivatives will cancel out. Unfortunately, this particular method always yields the trivial solution, so it is not very helpful. Indeed, we can imagine the directional derivatives in the tangent plane, and by symmetry they will always add to zero.

However, this leads us to try the next odd power of the directional derivative in the $y$-direction, $\frac{\partial^3}{\partial y^3}$.  Once again, the required rotational symmetry leads us to symmetrize the solution by adding to the third directional derivative in the $y$-direction the third directional derivative in the directions parallel to the other two sides. We can describe the resulting differential operator algebraically by rotating the directional derivative by $\sigma$:

\begin{align}
&\left(\nabla \cdot \hat{j}\right)^3 + \left(\nabla \cdot (\sigma \cdot \hat{j})\right)^3 + \left(\nabla \cdot (\sigma^2\cdot \hat{j})\right)^3 \nonumber\\
= & \left(\frac{\partial}{\partial y}\right)^3 + \frac{1}{8}\left(\sqrt{3}\frac{\partial}{\partial x}- \frac{\partial}{\partial y}\right)^3 + \frac{1}{8}\left(-\sqrt{3}\frac{\partial}{\partial x}- \frac{\partial}{\partial y}\right)^3 \nonumber \\
=& \frac{3}{4}\left(\frac{\partial^3}{\partial y^3} -3 \frac{\partial^3}{\partial y \partial x^2}\right)
\end{align}

This gives us our differential operator from Eq. (\ref{even_odd}), apart from a multiplicative constant (which is unimportant since the resulting solution will still need to be normalized). With this new understanding of Eq. (\ref{even_odd}), we can then argue as before that the new function $\hat{f_1}$ has the correct symmetry and satisfies the Dirichlet conditions. Using the Helmholtz relation, note that 
\begin{equation}
\frac{\partial^3}{\partial y \partial x^2} = - \frac{\partial^3}{\partial y^3}  - k^2 \frac{\partial}{\partial y}.
\end{equation}

\noindent Combining this with our differential operator, we can rewrite it as

\begin{equation}
 4 \frac{\partial^3}{\partial y^3}  +3 k^2 \frac{\partial}{\partial y}.
\end{equation}

\noindent Written in this way, it is clear the $\hat{f}_1$ is symmetric in the $x$-axis. The last thing we need to do is show that the solution is non-zero.  Without loss of generality we can assume there are no vertical nodal lines on the interior of the triangle. (This is possible using the last paragraph from Section \ref{sec:A1_A1}.)

Since any solution will be analytic on the interior of the triangle, we consider the power series of the function $f_2(x,y)$ at the origin.

\begin{equation}
f_2(x,y) = \sum_{i,j} c_{i,j}x^iy^j
\end{equation}

\noindent Let's consider the solution along the $y$-axis, $f_2(0,y)$. By symmetry, we know that there are only odd terms. Additionally, note that the directional derivatives at the origin parallel to each side are all zero, also by symmetry. We have previously argued that the sum of the three directional derivatives parallel to each side gives the zero function, thus $c_{0,1} = 0$.

\begin{equation}
f_2(0,y) = f_2(y) = \sum_{j=3,\text{ odd}} c_{0,j}y^j
\end{equation}

By assumption, $f_2(0,y)$ is not identically zero, so there is a non-zero term with minimal index. Now consider $\hat{f}_1$:

\begin{align}
\hat{f}_1 & = \left( 4 \frac{\partial^3}{\partial y^3}  +3 k^2 \frac{\partial}{\partial y}\right) f_2(y) \nonumber \\
& = 4 f_2'''(y)  +3 k^2 f_2'(y)
\end{align}

\noindent Using the first non-zero term in the power series for $f_2$, we note that its third derivative is non-zero, and higher order terms cannot cancel it. Thus $\hat{f}_1$ is non-zero, so we can normalize it to get a new solution $f_1$. In addition, we can note that $f_1$ is non-zero almost everywhere. Indeed, if $f_1$ is zero on an open set, then by analyticity it would be zero everywhere. 

While we have only shown that this process works for functions in A2 without a node along the $y$-axis, the strength of our conclusion shows that this is a local property. Combining this with earlier methods, the requirement to be non-zero along the $y$-axis can be eliminated.  

It should be noted that the only place we used the fact that $f_2$ was anti-symmetric was to prove that $f_1$ was non-zero. Indeed, the process introduced in Eq. (\ref{even_odd}) can also be used to transform a symmetric solution into an anti-symmetric solution, though the transformed solution may be zero. In the next section we will show that if a solution in class A1 is transformed by this differential operator and becomes the zero function, then it was (a harmonic of) the ``ground state." Otherwise, the transformed solution can be normalized to a solution in class A2, giving a one-to-one correspondence between solutions in class A2 and those solutions in A1 which are not harmonics of the ground state. 

\section{\label{sec:ground} The Ground State Solution}

In the previous section we introduced the differential operator defined in Eq. \ref{even_odd}) that transformed anti-symmetric solutions into symmetric solutions. One natural question is what does this operator do to the solution of the Helmholtz equation with the minimum value of $k^2$, that is, to the ``ground state solution''? Since the ground state solution is always non-degenerate, the operator in Eq. (\ref{even_odd}) must transform it into the zero function. We will now show that any solution from class A1 which transforms to zero under this differential operator is the ground state or a harmonic of the ground state. 

Let $f$ be a solution in symmetry class A1 which satisfies the two equations
\begin{equation}\label{gs1}
\left(\frac{\partial^2}{\partial x^2} + \frac{\partial^2}{\partial y^2} + k^2\right) f(x,y) =0
\end{equation}
and
\begin{equation}\label{gs2}
\left(\frac{\partial^3}{\partial y^3} - 3 \frac{\partial^3}{\partial y\partial x^2}\right) f(x,y) =0.
\end{equation}
Using the first equation we can eliminate partial derivatives with respect to $x$ from the second equation to obtain
\begin{align*}
0 & = \left(4 \frac{\partial^3}{\partial y^3} + 3 k^2 \frac{\partial}{\partial y}\right) f(x,y)\\
\implies 0& = \frac{\partial}{\partial y} \left(\frac{\partial^2}{\partial y^2} + \frac{3 k^2}{4} \right) f(x,y).
\end{align*}
Putting a solution of the form $f(x,y)= \sum X_i(x)Y_i(y)$ into the above equation we find that $Y_1 = 1$, $Y_2 = \sin(\frac{\sqrt{3}k}{2} y)$ and $Y_3 = \cos(\frac{\sqrt{3}k}{2} y)$. However, since solutions in class A1 are symmetric in $y$, the only allowed solutions are  $Y_1$ and $Y_3$. 

Similarly, putting $f(x,y)= \sum X_i(x)Y_i(y)$ into Eq.(\ref{gs2}), we get that $X_1 = A_1\cos(kx) + B_1\sin(kx)$ and $X_3 = A_3\cos(kx/2) + B_3\sin(kx/2)$. Imposing the boundary condition at $x=-\ell/2$, we find that $X_1(x) = \sin\left(k (x+\ell/2)\right)$ and $X_3(x) = \sin\left(k/2 (x+\ell/2)\right)$. Putting this all together we find the solution 

\begin{align*}
f(x) = &A \sin\left(k \Big(x+\frac{\ell}{2}\Big)\right) + B\sin\left(\frac{k}{2} \Big(x+\frac{\ell}{2}\Big)\right)\cos\left(\frac{\sqrt{3}k}{2} y\right).
\end{align*}

\noindent Imposing the remaining boundary condition along $y = \frac{-x+\ell}{\sqrt{3}}$, we get 
\begin{align*}
0 = & A \sin\left(k \Big(x+\frac{\ell}{2}\Big)\right) + B\sin\left(\frac{k}{2} \Big(x+\frac{\ell}{2}\Big)\right)\cos\left(\frac{\sqrt{3}k}{2} \left(\frac{-x+\ell}{\sqrt{3}}\right)\right) \\
& A \sin\left(k \Big(x+\frac{\ell}{2}\Big)\right) + B\sin\left(\frac{k}{2} \Big(x+\frac{\ell}{2}\Big)\right)\cos\left(\frac{k}{2} (x-\ell)\right) \\
& A \sin\left(k x+\frac{k\ell}{2}\Big)\right) + \frac{B}{2} \left(\sin\left(\frac{3 k \ell}{4}\right)+\sin\left(kx-\frac{k \ell}{4}\right)\right).
\end{align*}

\noindent This equation is satisfied when $B = \pm 2A$ and $\sin(3k\ell/4) = 0$, so $k = 4 \pi n/3 \ell$ for $n >0$. 
\begin{align*}
0 & = \sin\left(k x+\frac{k\ell}{2}\Big)\right) \pm \sin\left(kx-\frac{k \ell}{4}\right) \\
& = \sin\left(k x+\frac{2\pi n }{3}\Big)\right) \pm \sin\left(kx-\frac{\pi n}{3}\right) 
\end{align*}
Note that these two sine waves are horizontally shifted by $\pi n$, so there is a suitable choice of sign for $B$ to make them cancel. Putting this all together, 

\begin{align}\label{gs}
f_n(x) = &\sin\left(k_n \Big(x+\frac{\ell}{2}\Big)\right)+ 2(-1)^n  \sin\left(\frac{k_n}{2} \Big(x+\frac{\ell}{2}\Big)\right)\cos\left(\frac{\sqrt{3} \, k_n}{2} y\right),
\end{align}
where $k_n = \frac{4 \pi n}{3 \ell}$. For $n=1$ this agrees with the accepted solution in the literature for the ground state solution with center-to-vertex length $\ell$.\cite{Doncheski}  To compare Eq. (\ref{gs}) with the explicit solution given in McCartin, \cite{McCartin} note that if the inscribed circle has radius $r$ then $k= \frac{2 \pi}{3 r}$, and if the side length is $h$ then $k=\frac{4\pi \sqrt{3}}{3h}$. For larger values of $n$, we simply get the harmonics promised at the end of Section (\ref{sec:A2_A1}).

\section{\label{sec:con1}Conclusion}

In this paper we have examined the solutions to the Helmholtz equation $\nabla^2 \psi + k^2 \psi = 0$ within an equilateral triangle which obey the Dirichlet conditions on the boundary.  We have shown that every solution is a member of one of four symmetry classes and that, from symmetry considerations alone, any given solution in one symmetry class can be used to generate solutions in another symmetry class and/or with other values of the scalar $k^2$. We also used symmetry considerations to find a novel derivation of the ``ground state" solution of the Helmholtz equation.

These results have many interesting applications.  For example, in some cases we are looking for solutions to the Helmholtz equation which possess certain specific reflection or rotational symmetries.  Referring to the chart at the end of Section \ref{sec:rep3} we see that if we are looking for a solution with the symmetry properties of solutions in the symmetry class A2, we can generate such a solution if we already have another solution in either symmetry class A1 or symmetry class E2.  Similarly, for any given value of the scalar $k^2$, we can generate the ground state solution using Eq. (\ref{gs}) and then, since the ground state solution is in symmetry class A1, use the method presented in Section \ref{sec:rep3} to generate solutions with ``higher harmonics," that is, solutions whose scalar values are $n^2k^2$.  More generally, given any solution, we can generate many different solutions in different symmetries classes and with different values of the scalar $k^2$ from symmetry considerations alone, and keep generating solutions from each solution generated previously without ever having to solve the Helmholtz equation directly.

We end by noting that, when collected together, the techniques in this article shed light on the relationship between solutions within the $(30^{\circ},60^{\circ}, 90^{\circ})$ and equilateral triangles. For example, solutions to the equilateral triangle which are in symmetry classes A2 and E2 vanish along the $x$-axis. If we restrict our domain $\Delta$ to quadrants I and II, these solutions become solutions to the $(30^{\circ},60^{\circ}, 90^{\circ})$ triangle (with the same value of $k^2$). See Figures \ref{A2} and \ref{E2}, respectively.

Conversely, given a solution in the $(30^{\circ},60^{\circ}, 90^{\circ})$ triangle, we can reflect it across the $x$-axis to get a solution in the equilateral triangle that is in either of the symmetry classes A2 or E2. We can then take these solutions and construct from them solutions in A1 and E1 using the techniques from Section \ref{sec:A2_A1} and Section \ref{sec:rep5}, respectively. Since both of these techniques are reversible, we know that {\it all} solutions within the equilateral triangle can be found from the solutions in the $(30^{\circ},60^{\circ}, 90^{\circ})$ triangle except for the ground state solution and its harmonics.  However, these solutions can be derived directly using the technique in Section \ref{sec:A2_A1}, which means that knowing all of the solutions within the $(30^{\circ},60^{\circ}, 90^{\circ})$ triangle enables us to obtain all of the solutions within the equilateral triangle.  

In summary, apart from providing a new way to generate new solutions from any given solution to the Helmholtz equation within a triangle region, our method can also be used to establish two interesting connections between solutions to the Helmholtz equation within the equilateral and $(30^{\circ},60^{\circ}, 90^{\circ})$ triangles. First, the set of solutions within the $(30^{\circ},60^{\circ}, 90^{\circ})$ triangle is a subset of the set of solutions within the equilateral triangle. Second, we have a two-to-one correspondence between the solutions in the equilateral triangle which are not harmonics of the ground state and the solutions of the $(30^{\circ},60^{\circ}, 90^{\circ})$ triangle.

\section{Appendix I}\label{app1}

We want to show that the solution $g(x,y) = (f \circ T^{-1})(x,y) = f(X(x,y),Y(x,y))$ is a solution to the differential equation

$$\left(\frac{\partial^2}{\partial^2 x}+\frac{\partial^2}{\partial^2 y}\right) g(x,y) = 3 k^2 g(x,y)$$

\noindent when $f$ is a solution to the differential equation  

$$\left(\frac{\partial^2}{\partial^2 X}+\frac{\partial^2}{\partial^2 Y}\right) f(X,Y) = k^2 f(X,Y).$$

To make this a bit cleaner, we will need to write the transformation $T^{-1}(x,y)$ explicitly. Solving the system of equations

\begin{align*}
x & = \frac{1}{6} \left(3 X + \sqrt{3} Y + 3\right) \\
y & = \frac{1}{6} \left(\sqrt{3} X - 3 Y + \sqrt{3}\right)
\end{align*}

for $X$ and $Y$, we get

\begin{align*}
X(x,y) & = \frac{1}{4} \left(6 x - 2\sqrt{3} y + \sqrt{3}- 3\right) \\
Y(x,y) & = \frac{1}{4} \left(2\sqrt{3} x + 6 y - \sqrt{3} - 3 \right).
\end{align*}

In order to evaluate $\frac{\partial^2}{\partial^2 x} f(X,Y)$, we use the chain rule:

\begin{align*}
\frac{\partial^2}{\partial^2 x} f(X,Y)  = & \frac{\partial}{\partial x} \left(\frac{\partial X}{\partial x} \frac{\partial}{\partial X}f(X,Y) + \frac{\partial Y}{\partial x} \frac{\partial}{\partial Y}f(X,Y) \right)\\
= & \frac{\partial}{\partial x} \left(\frac{6}{4} \frac{\partial}{\partial X}f(X,Y) + \frac{2\sqrt{3}}{4} \frac{\partial}{\partial Y}f(X,Y) \right) \\
= & \frac{\partial}{\partial x} \left(\frac{3}{2} f_X(X,Y) + \frac{\sqrt{3}}{2} f_Y(X,Y) \right) \\
= & \frac{3}{2}  \frac{\partial}{\partial x} f_X(X,Y) + \frac{\sqrt{3}}{2} \frac{\partial}{\partial x} f_Y(X,Y) \\
= & \frac{3}{2}  \left(\frac{\partial X}{\partial x} \frac{\partial}{\partial X}f_X(X,Y) + \frac{\partial Y}{\partial x} \frac{\partial}{\partial Y}f_X (X,Y) \right) + \\
& \hspace{1cm} \frac{\sqrt{3}}{2} \left(\frac{\partial X}{\partial x} \frac{\partial}{\partial X}f_Y(X,Y) + \frac{\partial Y}{\partial x} \frac{\partial}{\partial Y}f_Y (X,Y) \right) \\
= & \frac{3}{2}  \left(\frac{3}{2} f_{XX}(X,Y) + \frac{\sqrt{3}}{2} f_{XY} (X,Y) \right) +  \\
& \hspace{1cm} \frac{\sqrt{3}}{2} \left(\frac{3}{2} f_{YX}(X,Y) + \frac{\sqrt{3}}{2} f_{YY} (X,Y) \right) \\
= & \frac{9}{4} f_{XX}(X,Y) + \frac{3\sqrt{3}}{2} f_{XY} (X,Y) + \frac{3}{4} f_{YY} (X,Y) 
\end{align*}

Similarly, for $\frac{\partial^2}{\partial^2 y} f(X,Y)$, we have

\begin{align*}
\frac{\partial^2}{\partial^2 y} f(X,Y)  = & \frac{\partial}{\partial y} \left(\frac{\partial X}{\partial y} \frac{\partial}{\partial X}f(X,Y) + \frac{\partial Y}{\partial y} \frac{\partial}{\partial Y}f(X,Y) \right)\\
= & \frac{\partial}{\partial x} \left(-\frac{2\sqrt{3}}{4} \frac{\partial}{\partial X}f(X,Y) + \frac{6}{4} \frac{\partial}{\partial Y}f(X,Y) \right) \\
= & \frac{\partial}{\partial y} \left(-\frac{\sqrt{3}}{2} f_X(X,Y) + \frac{3}{2} f_Y(X,Y) \right) \\
= & -\frac{\sqrt{3}}{2}  \frac{\partial}{\partial y} f_X(X,Y) + \frac{3}{2} \frac{\partial}{\partial y} f_Y(X,Y) \\
= & -\frac{\sqrt{3}}{2}  \left(\frac{\partial X}{\partial y} \frac{\partial}{\partial X}f_X(X,Y) + \frac{\partial Y}{\partial y} \frac{\partial}{\partial Y}f_X (X,Y) \right) + \\
& \hspace{1cm} \frac{3}{2} \left(\frac{\partial X}{\partial y} \frac{\partial}{\partial X}f_Y(X,Y) + \frac{\partial Y}{\partial y} \frac{\partial}{\partial Y}f_Y (X,Y) \right) \\
= & - \frac{\sqrt{3}}{2}  \left(-\frac{\sqrt{3}}{2} f_{XX}(X,Y) + \frac{3}{2} f_{XY} (X,Y) \right) +  \\
& \hspace{1cm} \frac{3}{2} \left(\frac{-\sqrt{3}}{2} f_{YX}(X,Y) + \frac{3}{2} f_{YY} (X,Y) \right) \\
= & \frac{3}{4} f_{XX}(X,Y) - \frac{3\sqrt{3}}{2} f_{XY} (X,Y) + \frac{9}{4} f_{YY} (X,Y) 
\end{align*}

Combining these two equations, we get

\begin{align*}
\nabla^2 g(x,y) = & \frac{\partial^2}{\partial^2 x} f(X,Y)+ \frac{\partial^2}{\partial^2 y} f(X,Y) \\
= &\frac{9}{4} f_{XX}(X,Y) + \frac{3\sqrt{3}}{2} f_{XY} (X,Y) + \frac{3}{4} f_{YY} (X,Y) \\
& \hspace{1cm} \frac{3}{4} f_{XX}(X,Y) - \frac{3\sqrt{3}}{2} f_{XY} (X,Y) + \frac{9}{4} f_{YY} (X,Y)  \\
= & 3 (f_{XX}(X,Y) + f_{YY}(X,Y)) \\
= & 3 k^2 f(X,Y) \\
= & 3 k^2 g(x,y) 
\end{align*}

which is the desired result.
\section{Appendix II}\label{app2}

In this section we examine the properties of solutions obtained by de-composing the fundamental domain into $n^2$ equilateral triangles.  First we show that if we start with a solution $f$ in A2 or E2 which is a solution of the Helmholtz equation with the scalar $k^2$ then, by decomposing the triangular domain into $n^2$ equilateral triangles, we can generate a solution to the Helmholtz equation with scalar $n^2k^2$. 

The transformation of the plane which replaces the original triangle by $n^2$ triangles is a pure dilation by a factor of $n$, followed by a translation along the $x$ axis. The translation does not affect the scalar in Helmholtz's equation, so we will just call it $C$. Thus our transformation is

\begin{align*}
X(x,y) & = n x + C \\
Y(x,y) & = n y.
\end{align*}

An easier way to see the affect such a transformation would have on the scalar, consider the Jacobian matrix for the transformation

\begin{equation*}
\mathcal{J} = \begin{pmatrix} n & 0 \\ 0 & n \end{pmatrix}.
\end{equation*}

The determinant of this matrix is $n^2$, and gives us the desired result.

Note that if the original solution has rotational symmetry (i.e. is in the symmetry class A2), then the transformation will not change the symmetry class.   This is easy to see since, when the equilateral triangle is subdivided into $n^2$ triangles, the resulting picture is rotationally symmetric. Since the solution inside each of these smaller triangles is the same rotationally symmetric symmetric solution, the resulting solution is rotationally symmetric.

However, if it started asymmetric (i.e. in the symmetry class E2) then this approach will lead to a solution in A2 iff $n$ is divisible by 3.  As noted above, the subdivided triangle is rotationally symmetric. However, since we are starting with a solution in E2, the resulting picture may or may not be rotationally symmetric. To test for rotational symmetry, consider Figure \ref{a9}.

Looking at the solution inside the small triangle at the left of the subdivided triangle, we can sketch in the solution on the large subdivided triangle by reflecting this small triangle over the boundary and keep track of the nodal line. Comparing the small triangles along the top edge, we can see that after two reflections, the solutions appears rotated by $\rho^{-1}$.

Restricting ourselves to looking only at the 3 small triangles at the tips of a subdivided triangle, note that a rotation of the large triangle induces a rotation of the small triangles. This rotation only agrees with the reflection method if $\rho = \rho^{-(n-1)}$. Equivalently, $\rho^n = \mathbf{1}$, which means that $n$ must be a multiple of $3$.

Given a solution with E2 symmetry and scalar $k^2$, the harmonic with scalar $3^2k^2$ is the same as the solution resulting from using the method from Section \ref{sec:E2_A2}) twice $k^2 \rightarrow 3 k^2 \rightarrow 9 k^2$. It is easiest to see this by comparing the two pictures, shown in Figure \ref{bst_vs}. On the left, we have the picture using the harmonic method, and on the right we have the picture resulting from using the method from Section \ref{sec:E2_A2}) twice. Following the two nodal patterns, we see that they are the same.

\begin{acknowledgments}
We would like to thank Peter Wong and Matthew Cot\'{e} for many helpful discussions, and for advising the senior thesis \cite{Stambaugh} on which much of the work presented here is based.
\end{acknowledgments}

\end{document}